 \documentclass[manuscript ]{aastex631}

\shorttitle{Strong electron-dominated current sheets in the solar wind}
\shortauthors{Khabarova et al.}

\begin{document}

\title{Electron-to-ion bulk speed ratio as a parameter reflecting the occurrence of strong electron-dominated current sheets in the solar wind \\ 
ACCEPTED FOR PUBLICATION IN THE ASTROPHYSICAL JOURNAL}

\author[0000-0002-5700-987X]{Olga Khabarova}
\affiliation{Pushkov Institute of Terrestrial Magnetism, Ionosphere and Radio Wave Propagation of the Russian Academy of Sciences (IZMIRAN), 
108840 Moscow, Russia; habarova@izmiran.ru}
\affiliation{Space Research Institute of the Russian Academy of Sciences (IKI RAS), 117997 Moscow, Russia}

\author[0000-0002-5700-987X]{J\"org~B\"uchner}
\affiliation{Max Planck Institute for Solar System Research, 37077 G\"ottingen, Germany; buechner@mps.mpg.de}
\affiliation{Center for Astronomy and Astrophysics, Technical University of Berlin, 10623 Berlin, Germany}

\author[0000-0002-5700-987X]{Neeraj Jain}
\affiliation{Center for Astronomy and Astrophysics, Technical University of Berlin, 10623 Berlin, Germany}

\author[0000-0002-5700-987X]{Timothy Sagitov}
\affiliation{Pushkov Institute of Terrestrial Magnetism, Ionosphere and Radio Wave Propagation of the Russian Academy of Sciences (IZMIRAN), 
108840 Moscow, Russia; habarova@izmiran.ru}

\author[0000-0002-5700-987X]{Helmi Malova}
\affiliation{Space Research Institute of the Russian Academy of Sciences (IKI RAS), 117997 Moscow, Russia}
\affiliation{Scobeltsyn Nuclear Physics Institute (NPI), Lomonosov State University, 119991 Moscow, Russia}

\author[0000-0002-4319-8083]{Roman Kislov}
\affiliation{Space Research Institute of the Russian Academy of Sciences (IKI RAS), 117997 Moscow, Russia}

\begin{abstract}
{
Current sheets (CSs) are preferred sites of magnetic reconnection and energy dissipation in collisionless 
astrophysical plasmas. Electric currents in them may be carried by both electrons and ions. In our prior theoretical studies of processes associated with the CS formation in turbulent plasmas,
for which we utilized fully kinetic and hybrid code simulations with ions considered as particles and electrons - as a massless fluid ~\citep{jain2020}, we found that electron-dominated CSs may form inside or nearby ion-dominated CSs ~\citep{azizabadi2021}. Electrons become the main carrier of the electric current and contributors to energy dissipation in electron-dominated CSs. These magneto-plasma structures represent a distinguished type of CSs and should not be mixed up with so-called electron-scale CSs. Our simulations show that such CSs are characterized by the electron-to-ion bulk speed ratio $u_e/u_i$ increases. Theoretical predictions and high-resolution observations from the MMS mission suggest that strong electron-dominated CSs can be seen at ion scales. Therefore we suggest that applying the $u_e/u_i$ parameter to the solar wind data may allow locating the strongest electron-dominated CSs with an ordinary spacecraft resolution of one-three seconds at least approximately. The results show that, indeed, an impact of electron-dominated CSs on the plasma observed during a period of quiet solar wind conditions at 1 AU may be associated with sharp changes in $u_e/u_i$. Electron-dominated CSs are found to be localized in the vicinity of ion-dominated CSs identified via changes in the magnetic field and plasma parameters ~\citep{khabarova2021}, displaying 
the same clustering. 
We conclude that $u_e/u_i$ may be used as one of key parameters for 
probing CSs in the solar wind and the role of electrons in them.
}
\end{abstract}



\section{Introduction}
\label{sec:introduction}

Spacecraft routinely observe  electric-current-carrying thin plasma layers, current sheets (CSs), in collisionless space plasmas (e.g., ~\citep{Nakamura2006,sundkvist2007, Greco2009,podesta2017,azizabadi2021,jain2020,khabarova2021}). In the solar wind, CSs are formed at discontinuities that separate regions with differently-directed magnetic fields ~\citep{Syrovatski1971}. Such discontinuities may represent a continuation of large-scale neutral lines of the solar origin as well as form at edges of various streams and flows, between magnetic islands and result from magnetic reconnection, instablities and wave propagation ~\citep{Khabarova2021SSRv}. CS structures are known to play a significant role in the development of turbulence and energy release in a form of heating or particle acceleration  ~\citep{MunozKilianBuchner2014,Khabarova2015,Khabarova2016,MunozBuechner2018,jain2020,Lazarian2020,azizabadi2021,Khabarova2021SSRv,pezzi2021, Pezzi2022}. 
They can  contribute to an energy cascade when the magnetic energy is transported from larger to shorter 
scales until it is transferred to the kinetic energy of particles via magnetic reconnection and/or dissipation.

In collisionless plasmas, CSs may thin down to kinetic plasma scales, such as the inertial length or gyro-radii of particles, whichever is reached 
earlier~\citep{jain2020,azizabadi2021}.
Then kinetic instabilities would cause additional, small scale turbulence which directly dissipates energy or allows fast magnetic 
reconnection. 
This sequence of events is well-known for large, long-lived CSs of the solar origin, such as the 
heliospheric current sheet (HCS). Dynamical processes occurring at the HCS create a wide cloud of secondary, smaller-scale CSs and other dynamically 
evolving coherent structures in its vicinity (see ~\citep{Khabarova2021SSRv} and references therein).
The specific of dissipation mechanisms, a threshold of micro-instabilities and an efficiency of the energy conversion depend on the structure and properties of CSs, in particular, on the kind of particles carrying the electric current, their possible anisotropic distribution and other macro- and microscopic plasma parameters.

 On the other hand, turbulence can create thin and short-lived CSs (e.g.,~\citep{howes2016}) 
as found in numerous numerical 
simulations of dynamical processes in space turbulent plasmas ~\citep{maron2001,franci2015,perri2012,howes2016, Pezzi2022}. In the solar wind, this scenario realizes far from long-lived and large-scale CSs, in undisturbed plasma. CSs created by turbulence may merge and form larger and longer-lived structures if plasma is impacted by waves, instabilities or flows.

In order to understand peculiarities of the CS formation, mainly macroscopic, fluid-type numerical simulations have been carried out (see, e.g. ~\citep{Biskamp1989,Barta2010,Barta2011ApJ1}). 
Based on the results of restricted electron-MHD simulations, it has been suggested that the CSs with a thicknesses ranging 
down to electron scales are responsible for structuring 3D magnetic reconnection ~\citep{JainBuchner2014-1,JainBuchner2014-2}. 
Both observational studies and numerical simulations suggest that a large fraction of the total magnetic energy is dissipated in and around kinetic-scale-CSs that form self-consistently and possess a significant power of turbulence ~\citep{borovsky2010, matthaeus2015}. 

{  
The occurrence of CSs is known to determine the shape of the power spectrum of magnetic field variations in the solar wind. This interesting fact has been discovered by Gang Li in 2011 ~\citep{Lietal2011} and then recently confirmed by Borovsky and Burkholder ~\citep{BorovskyBurkholder2020}. Gang Li showed that the current-sheet-abundant solar wind is characterized by the Kolmogorov-like power spectrum with the slope of -1.7, and the solar wind without CSs demonstrates Iroshnikov-Kraichnan scaling with a slope of -1.5. Borovsky and Burkholder ~\citep{BorovskyBurkholder2020} performed an analysis of factors forming a shape of the spectrum and concluded that both purely topological characteristics of CSs and dynamical processes occurring in them and their vicinity impact the spectrum considerably.

Since a dissipation mechanism in and around CSs is not quite clear yet, their visual or automated identification in space and subsequent thorough statistical studies of their properties are crucial to understand numerous processes associated with these plasma objects. 
Satellite observations in the Earth’s magnetosphere, spacecraft observations in the solar wind and theoretical investigations allowed understanding general properties of proton-current-dominated CSs with a width up to several proton gyroradii. As for electron, thinner currents, prior limitations of thin CS models and insufficiency of spatial/temporal resolution of observations gave only rough estimates of the thickness and the amplitude of the electron current peak (see ~\citep{Zelenyi_et_al2004,Zelenyi_et_al2011, Zelenyi_et_al2019,Malova_et_al2012,Malova_et_al2013, Malova_et_al2017} and references therein). 

It must be stressed that the term "electron CS" sometimes used in the literature may be confusing because this mixes up electron-scale CSs with a width of several electron gyroradii and electron-dominated CSs of any origin in which the electric current is mostly carried by electrons. Sometimes this is the same since CSs in which electrons carry the current can be very thin. However, the term "electron-dominated CS" has a wider meaning because it does not impose a thickness limit on the particular CS.  Theoretical estimations show that the width of electron currents in the solar wind CSs can be of one-two ion giroradii ~\citep{ Malova_et_al2017}, and one may expect that very strong electron currents in electron-dominated CSs and the effects in plasma may even be wider ~\citep{Pezzi2021mnras, Vasko2021}. The current study considers not electron-scale CSs but electron-dominated CSs. We will show below that strong electron-dominated CSs can be as wide as ion CSs.

An appearance of modern instruments with a high resolution for the magnetic field have provided researchers an opportunity to investigate a new kind of CSs called super-thin CSs (STCSs) in magnetospheres. Such CSs have been observed in the course of new spacecraft missions such as MAVEN in the Martian magnetotail ~\citep{Grigorenko_et_al2019}. Recent studies confirm the existence of STCSs in the terrestrial magnetotail ~\citep{Leonenko_et_al2021}. A half-thickness of STCSs is about a few or less electron gyroradii, therefore, these current layers can be considered as electron-dominated very thin CSs. The Magnetospheric Multiscale (MMS) mission with its electron-scale tetrahedron configuration has been very useful in understanding properties of electron-scale STCSs in both the Earth’s magnetopause and the magnetotail (e.g., ~\citep{Phan_et_al2016,Dong_et_al2018, Leonenko_et_al2021}).

Contrary to the magnetosphere, there have not been investigations of properties of CSs determined by the electron currents in the heliosphere so far. Studies of CSs in the solar wind are focused on ion CSs only.  Various criteria are used for identifying proton- or ion-dominated-CSs. Most commonly, significant rotation of the magnetic field vector (sometimes, from one direction to the opposite) and signatures of the crossing of a neutral line are employed to distinguish between ordinary discontinuities and CSs. These are primary signs of CS crossings. Additionally, observers analyze the behavior of plasma parameters, namely, the plasma beta ($\beta$) that usually sharply increases at strong ion CSs and the ratio of the Alfv\'en speed to the solar wind speed (that decreases at CSs). These are secondary signatures identifying ion-dominated CSs. An overview of both visual and automated methods of CS identification in the heliosphere can be found in ~\citep{khabarova2021}.

An identification of thin current layers of electron scales based on data from most spacecraft operating in the solar wind is still complicated (e.g., ~\citep{Kellogg_et_al2003,Kellogg_et_al2006}). The main reason for that is the insufficiently of the resolution of measurements. Spacecraft typically allow only one-three second measurements of plasma parameters and the magnetic field which far larger than electron scales. The second reason is the absence of multi-scale missions owing to which thin electron-scale CSs were discovered and studied in the terrestrial magnetotail ~\citep{Sergeev_et_al1993, Runov_et_al2003,Runov_et_al2006}. All measurements in the solar wind are single-spacecraft. The only mission that could help is MMS since a part of its orbit lies in the solar wind. However, MMS leaves the terrestrial magnetosphere and stay in the solar wind only for a short time during which the burst mode is mostly switched off. This does not allow performing comprehensive studies of thin electron-scale CSs in the solar wind.

Meanwhile, since strong electron-dominated CSs in the solar wind can be wider than electron-scale CSs, the task of their finding outside the terrestrial magnetosphere does not seem hopeless. Below, we will show an example of crossing of such a CS with MMS in the solar wind. If one knows specific features characterizing the occurrence of strong electron-dominated CSs in the solar wind, then it is possible to apply commonly accepted techniques for CS identifying to recognition of electron-dominated CSs from the in situ data, using an ordinary spacecraft resolution of one-three seconds ~\citep{khabarova2021}. It also would be interesting and useful to find secondary or indirect signatures of strong electron-carried currents. 

Theoretical studies and numerical simulations may help work out a problem. A theoretical approach in a frame of a hybrid 1D or 2D models in which ions are considered with a quasi-adiabatic approach and the electron motion is treated as an MHD flow is known to be the most perspective in the description of such thin multilayered CSs ~\citep{Zelenyi_et_al2004,Zelenyi_et_al2011, Petrukovich_et_al2011,Malova_et_al2012,Malova_et_al2013}. A comprehensive work ~\citep{Zelenyi_et_al2020} has provided the basis of the theory of STCSs, and recent numerical simulations allowed finding the way to identify electron-dominated CSs via an analysis of spatial variations of plasma parameters  ~\citep{azizabadi2021, jain2020}.}

For a better understanding of the CS formation and their expected thinning 
down to kinetic scales different kinds of numerical simulations have been 
carried out utilizing a variety of different plasma models like, e.g., 
hybrid codes which consider ions as particles and electrons as a 
fluid ~\citep{azizabadi2021, jain2020}.
The latter investigation revealed, supported by theoretical estimates, 
an extra-criterion which can be used for a better understanding of the 
structure of CSs in turbulent plasmas.
This criterion is based on the finding that within thinning CSs the (shear) 
flow velocity of the current carrying electrons in the direction 
parallel to the ambient magnetic field  ($|u_e|$) should  significantly exceed by 
large the ion bulk flow velocity ($|u_i|$) in this direction.
At the same time the plasma density would vary only weakly 
(less than $10\%$) throughout the CSs. Thus, the ratio of 
electron over the ion bulk flow velocities ($|u_e/u_i|$) should 
become very large at the strongest electron-dominated CSs in the solar wind.

Theoretical studies predict that ion- and electron-dominated CSs may be observed by a spececraft in different ways. First, a thin electron-dominated CS can be embedded in the wider ion-dominated CS ~\citep{Malova_et_al2017}. Second, spatially-separated electric currents in which electrons and ions are main carries may form at reconnecting CSs, owing to the effect of partial separation of charges ~\citep{Zharkova&Khabarova2012, Zharkova&Khabarova2015, Khabarova2020}. In that case they may be observed with a high resolution as two closely located CSs dominated by electrons and ions, but in fact these are two parts of the same bifurcated CS with the currents spatially separated with respect to the main neutral plane. It is not clear if closely lying ion- and electron-dominated CSs can be found from in situ observations via variations in plasma parameters in this case. There are no predictions regarding formation of totally isolated (single) ion- or electron-dominated CSs in the solar wind either, and we know nothing about their possible survival time if such CSs form due to some non-stationary processes. No one knows if they may live for the time period sufficient for their observation with spacecraft. Therefore, the existence of electron-dominated CSs raises a lot of questions, and it would be important to carry out a study that allowed finding an approximate location of electron-dominated CSs using their known impact on plasma parameters.   

The structure of the manuscript is as follows. 
We first carry out simulations that show formation of CSs in a turbulent plasma and describe the jump criterion of $|u_e/u_i|$ 
within CSs from the theoretical point of view (see section~\ref{sec:simulations} and section~\ref{sec:theory}, respectively). Then we mimic a crossing of several simulated CSs with a virtual spacecraft to show that the corresponding sharp variations in $|u_e/u_i|$ can potentially be spotted by a spacecraft in the solar wind with a typical resolution of one-three seconds (section~\ref{sec:theory}).
In section~\ref{sec:data} we describe the observational approach to the problem of finding the impact of electron currents on the ambient plasma reflected in variations of plasma parameters. A supporting case study of the CS crossing in the solar wind with MMS follows the theoretical part. It is aimed at a preliminary estimation of the ability of the $u_e/u_i$ parameter to recognize plasma structures associated with electron-dominated CSs. Then we utilize the WIND spacecraft observations identifying such magneto-plasma structures and comparing their found locations with those of ion-dominated CSs found, using the method described by ~\citep{khabarova2021}. Finding an approximate location of electron-dominated CSs may be very useful for future statistical studies of solar wind CSs in which electrons are the main current carriers. The conclusions of the investigation are drawn and discussed in 
section \ref{sec:conclusion}.    

\section{Simulation results}
\label{sec:simulations}

We carried out hybrid code simulations of a turbulent plasma in which
we treated ions as particles and electrons as an inertia-less fluid on a 
two-dimensional mesh spanning over an x-y plane.
For this sake we utilized the PIC-hybrid code A.I.K.E.F. ~\citep{mueller2011}. 
We initialized the simulations with random-phased fluctuating magnetic 
fields and plasma velocities within a wave number range
$|k_{x,y}d_i| < 0.2$ ($k_{x,y}\neq 0$). 
Here $k_x$ and $k_y$ are wave numbers in the x- and y-directions, 
respectively, 
$d_i=v_{Ai}/\omega_{ci}$, $v_{Ai}=B_0/\sqrt{\mu_0n_0m_i}$ and 
$\omega_{ci}=e B_0/ m_i$  are inertial length, Alfv\'en velocity and 
cyclotron frequency of ions, respectively ($\mu_0$ is the vacuum magnetic  
permeability, $e$ the electron charge and $m_i$ the proton mass).
The fluctuations are  imposed on an isotropic background plasma 
of uniform density $n_0$.  
All initialized modes have the same energy and a root-mean-square 
value $B_{rms}/B_0=0.24$, where $B_0$ is the uniform magnetic 
field applied perpendicular to the simulation plane. 
Electron and ion plasma beta are $\beta_e=2\mu_0 n_0 k_B T_e/B_0^2=0.5$ 
and $\beta_i=2\mu_0 n_0 k_B T_i/B_0^2=0.5$, 
with $T_e$ and $T_i$  being the electron and ion 
temperatures, respectively and 
$k_B$ the Botzmann constant. 
The simulation box size  $256 d_i \times 256 d_i$
is resolved  by $512 \times 512$ grid points with 500 macro-particles 
per cell. 
The time step was chosen to be $\Delta t$=0.01 $\omega_{ci}^{-1}$.  
Periodic boundary conditions are applied in all directions. 

\begin{figure}
  \begin{center}
    \includegraphics[clip=true,trim=0.25cm 0.7cm 0.25cm 0.7cm,width=0.49\linewidth]{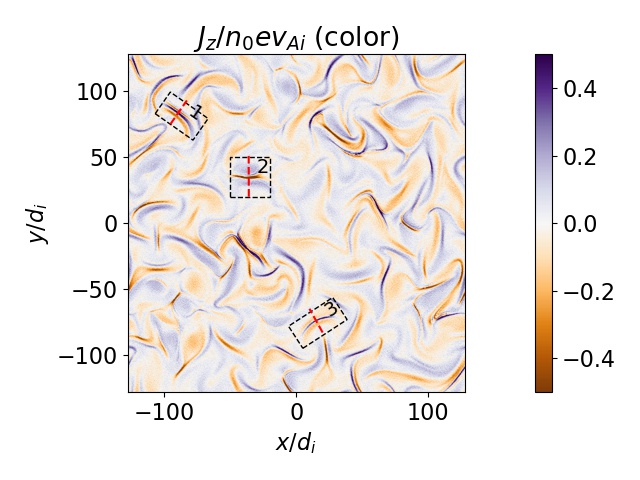}
    \thicklines
    \put(-250,27){\vector(1,0){20}}
    \put(-250,27){\vector(0,1){20}}
    \put(-250,27){\vector(-1,-1){12}}
    \put(-230,20){\large{\bf x}}
    \put(-250,50){\large{\bf y}}
    \put(-270,20){\large{\bf z}}
    \put(-260,5){\large{\bf $B_0$}}
        \includegraphics[clip=true,trim=0.25cm 0.7cm 0.25cm 0.7cm,width=0.49\linewidth]{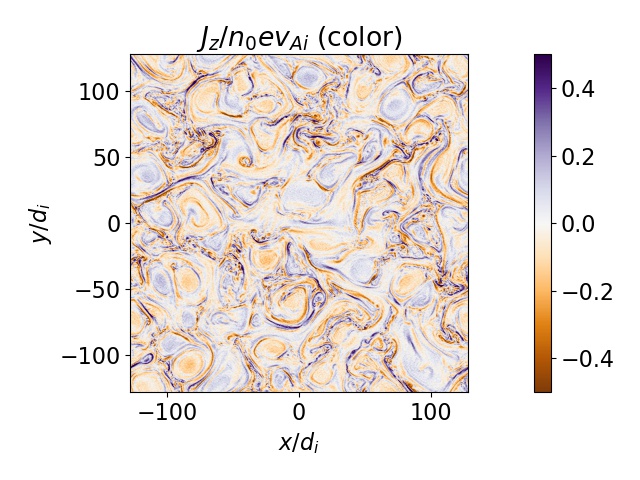}
        \caption{Parallel current density $J_z$  at two moments of time:
        $\omega_{ci}t=50$ (left column) and $\omega_{ci}t=150$ (right column). Three CSs, numbered 1, 2 and 3,  are highlighted at $\omega_{ci}t=50$ by enclosing them in rectangles with dashed borders. The red line in each rectangle is the current sheet normal. Mean magnetic field is in the out-of-plane ($z$) direction as shown by arrows on the left.}
        \label{fig:jz}
    \end{center}
\end{figure}

In a course of the evolution of the initially  long-wavelength magnetic and ion velocity fluctuations,  
CSs are formed by $\omega_{ci}t=50$ (Fig. \ref{fig:jz}).
These CSs later break up developing 
shorter wavelength 
turbulence ~\citep{daughton2011,Munoz2018,dahlin2015}, 
as shown at $\omega_{ci}t=150$ in Fig. \ref{fig:jz}. 
Our hybrid code simulations revealed that within the CSs the parallel (to $B_0$) electron bulk flow velocities became much larger than the parallel ion bulk velocities ~\citep{jain2020}. The perpendicular (to $B_0$) bulk velocities of electrons and ions, on the other hand, are of the same order but smaller than the parallel bulk velocity of electrons.
Therefore net bulk speed of electrons is larger than that of ions.  

\begin{figure}
  \begin{center}
    \includegraphics[clip=true,trim=0.25cm 0.7cm 0.25cm 0.7cm,width=0.49\linewidth]{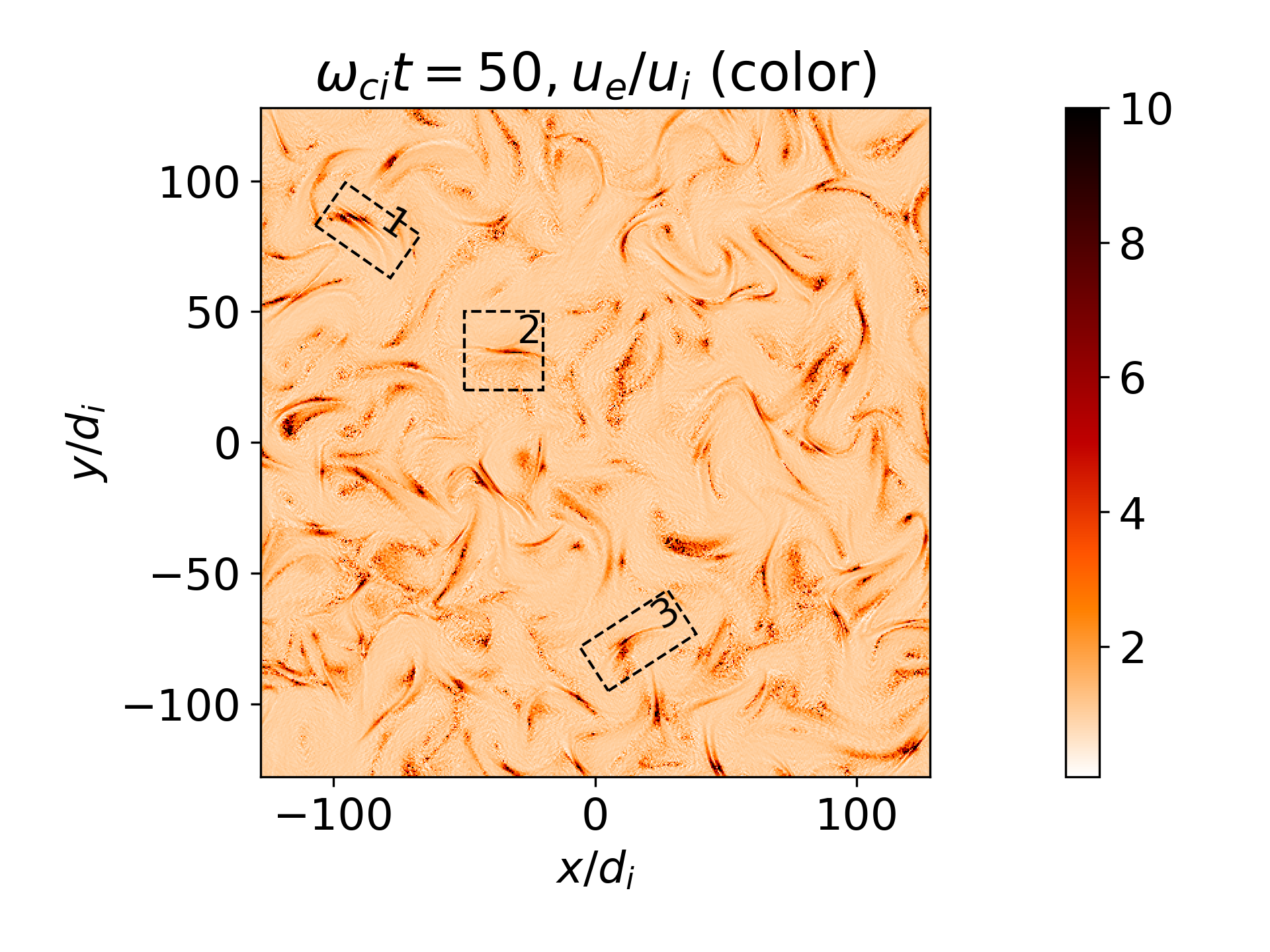}
        \includegraphics[clip=true,trim=0.25cm 0.7cm 0.25cm 0.7cm,width=0.49\linewidth]{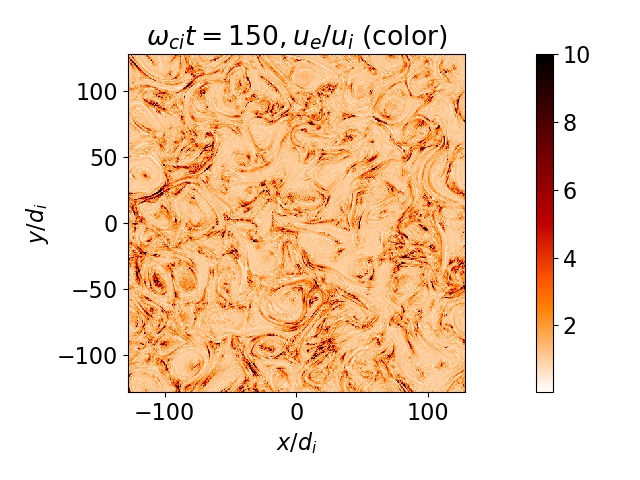}
        \caption{Ratio $u_e/u_i$ of the magnitudes of the electron- ($u_e=|\mathbf{u_e}|$) to the 
        ion-bulk speeds ($u_i=|\mathbf{u_i}|$) at two moments of time:
        $\omega_{ci}t=50$ (left column) and $\omega_{ci}t=150$ (right column). Rectangles with dashed border enclose the three current sheets (CS-1, CS-2 and CS-3) at $\omega_{ci}t=50$.}
        \label{fig:ueDui}
    \end{center}
\end{figure}

In space observations, special care has to be taken to distinguish 
between parallel and perpendicular (to the magnetic field) 
velocity components. It is typically easier to get net velocity of particles in space observations.  
In order to interpret the  observations we examined the ratio $u_e/u_i$ of the net bulk velocities 
$u_e = |\mathbf{u}_{e}|$ and $u_i = |\mathbf{u}_{i}|$ of the electrons and ions, 
respectively. 
As an example Fig.~\ref{fig:ueDui} depicts the isolines
of $|\mathbf{u}_{e}|/|\mathbf{u}_{i}|$ in the simulation plane at 
$\omega_{ci}t=50$ and $\omega_{ci}t=150$.
It can be seen that the electron bulk speed $u_{e}$ exceeds the ion bulk speed $u_i$ by several 
times in ion-scale CSs. 
Moreover, the ratio $u_e/u_i$ enhances as the turbulence evolves from $\omega_{ci}t=50$ to 150. 

\begin{figure}
  \begin{center}
    \includegraphics[clip=true,trim=0.25cm 0.7cm 0.25cm 0.7cm,width=0.49\linewidth]{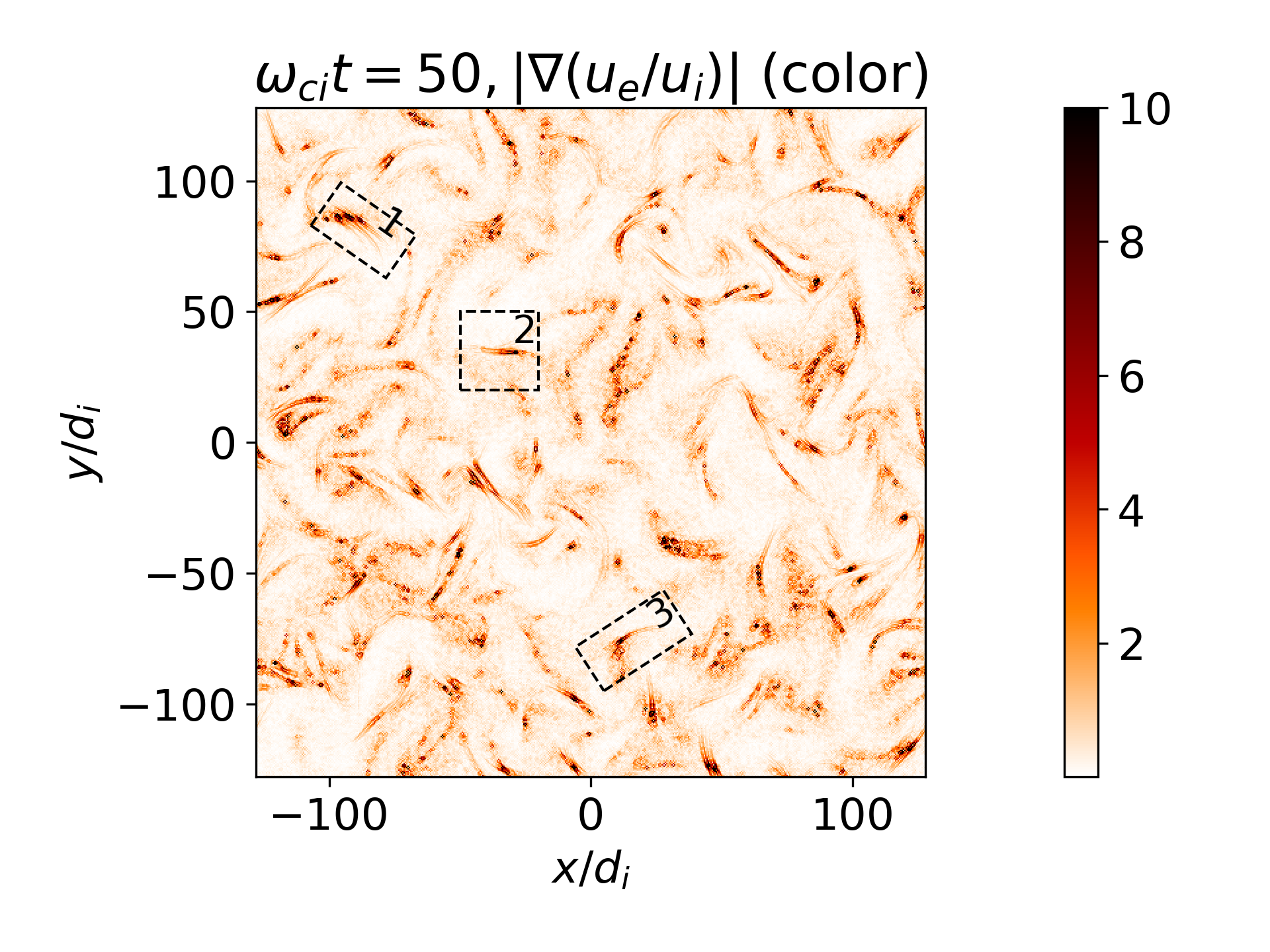}
        \includegraphics[clip=true,trim=0.25cm 0.7cm 0.25cm 0.7cm,width=0.49\linewidth]{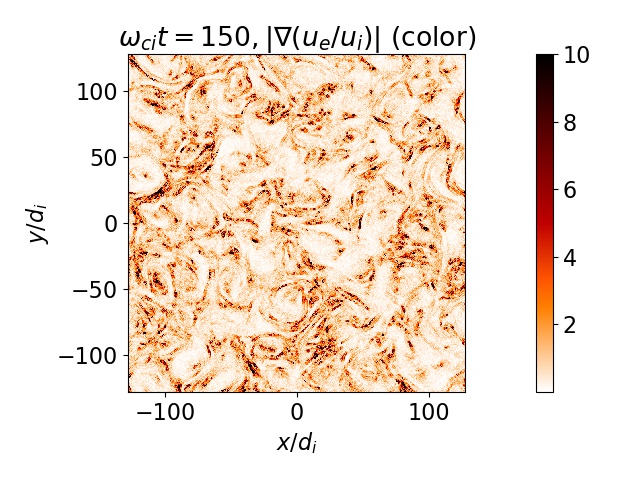}
        \caption{Magnitude of the gradient of the ratio $u_e/u_i$ at two moments of time:
        $\omega_{ci}t=50$ (left column) and $\omega_{ci}t=150$ (right column).  Rectangles with dashed border enclose the three current sheets (CS-1, CS-2 and CS-3) at $\omega_{ci}t=50$.}
        \label{fig:gradient_ueDui}
    \end{center}
\end{figure}

In order to detect a jump in $u_e/u_i$ in time series measurements by spacecraft, it is more practical to look at the derivative of $u_e/u_i$. 
Figure \ref{fig:gradient_ueDui} shows the magnitude of the spatial gradient of $u_e/u_i$ in the simulation plane. It is clear from Figs. \ref{fig:ueDui} and \ref{fig:gradient_ueDui} that the value of $|\nabla(u_e/u_i)|$ 
in CSs is better distinguished from its value outside CSs as compared to the values of $u_e/u_i$.

\begin{figure}
  \begin{center}
    \includegraphics[clip=true,trim=0.25cm 0.7cm 0.25cm 0.7cm,width=0.32\linewidth]{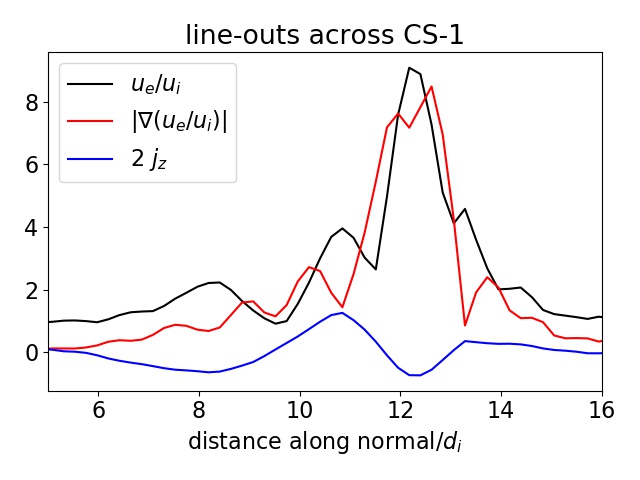}
    \includegraphics[clip=true,trim=0.25cm 0.7cm 0.25cm 0.7cm,width=0.32\linewidth]{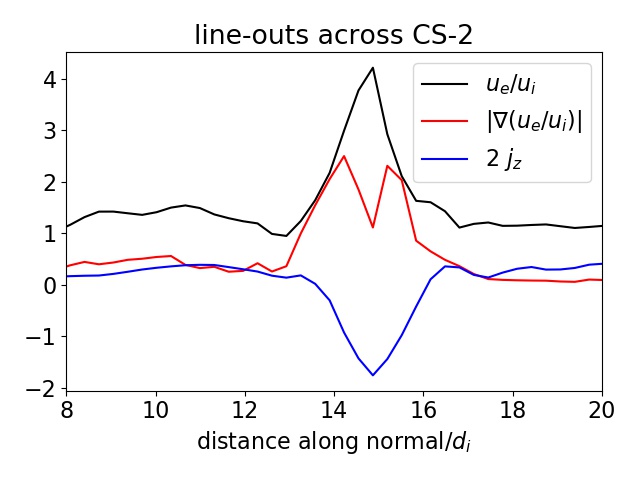}
    \includegraphics[clip=true,trim=0.25cm 0.7cm 0.25cm 0.7cm,width=0.32\linewidth]{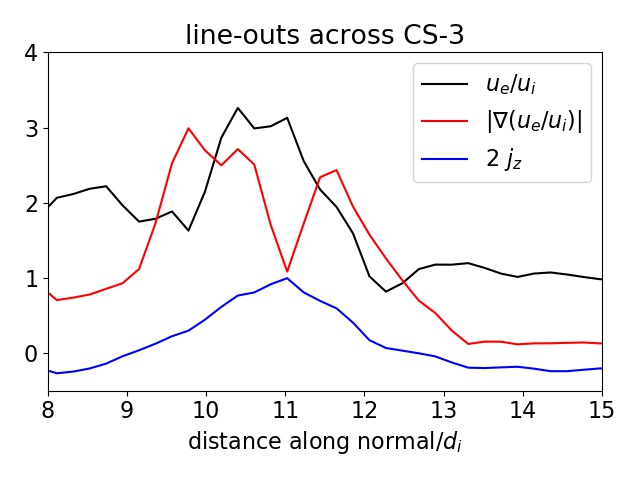}
        \caption{Line-outs of $u_e/u_i$, $\nabla(u_e/u_i)$ and $j_z$ across the three CSs (CS-1, CS-2, CS-3) highlighted in Fig. \ref{fig:jz} at $\omega_{ci}t=50$.}
        \label{fig:lineouts_ueDui}
    \end{center}
\end{figure}

Fig. \ref{fig:lineouts_ueDui} shows the line-outs of $u_e/u_i$, $|\nabla(u_e/u_i)|$ and $J_z$ along the normal of the three CSs (CS-1, CS-2 and CS-3) highlighted in Fig. \ref{fig:jz}.  It can be seen that $u_e/u_i$ takes a jump from its value of the order of unity outside CSs to at least several times larger value 
in CSs. 
Note that the value of $u_e/u_i$ inside CSs sheets is not unique. 
It might be different for different CSs. 
Therefore the actual value of $u_e/u_i$ inside CSs is not as important as 
the jump in its value from outside to inside the sheets as far as the CS
detection is concerned. 
This jump is characterized by $|\nabla (u_e/u_i)|$. 
Note that $\nabla(u_e/u_i)$ inside CSs is dominated by the gradient along 
the current sheet normal which changes sign across the current sheet. 
A dip in the value of $|\nabla(u_e/u_i)|$ at the peak of $u_e/u_i$ in 
Fig. \ref{fig:gradient_ueDui} corresponds to this change of sign of $\nabla(u_e/u_i)$. 

\begin{figure}
  \begin{center}
    \includegraphics[width=\linewidth]{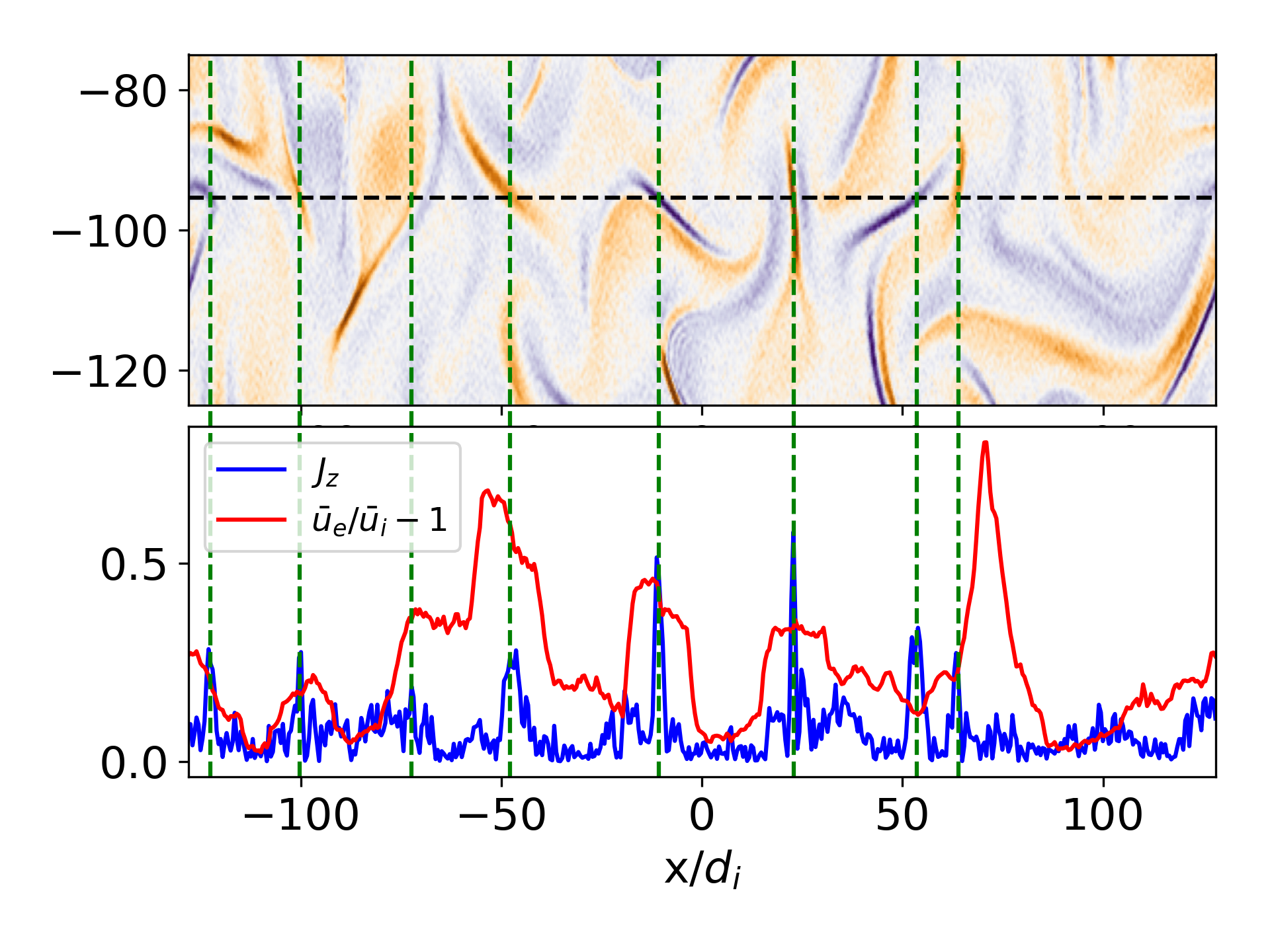}
        \caption{Color coded $j_z$ at $\omega_{ci}t=50$ in a subdomain ($y \in [-125,-75]$ on the vertical axis and $x \in [-128,128]$ on the horizontal axis) of the x-y simulation plane (top panel). The horizontal dashed line drawn at $y\approx -95.4$ in the top panel is an assumed spacecraft trajectory.  Line-outs of $j_z$ and  $\bar{u}_e/\bar{u}_i-1$ along the assumed spacecraft trajectory (bottom panel). Here $\bar{u}_e$ and $\bar{u}_i$ are the values of $u_e$ and $u_i$ averaged over a distance of 16 $d_i$ at each point on the trajectory, respectively. Green vertical lines cross the trajectory at locations of CSs.}
        \label{fig:trajectory_lineouts_ueDui}
    \end{center}
\end{figure}

\begin{figure}
  \begin{center}
    \includegraphics[clip=true,trim=0.25cm 0.7cm 0.25cm 2.1cm,width=0.32\linewidth]{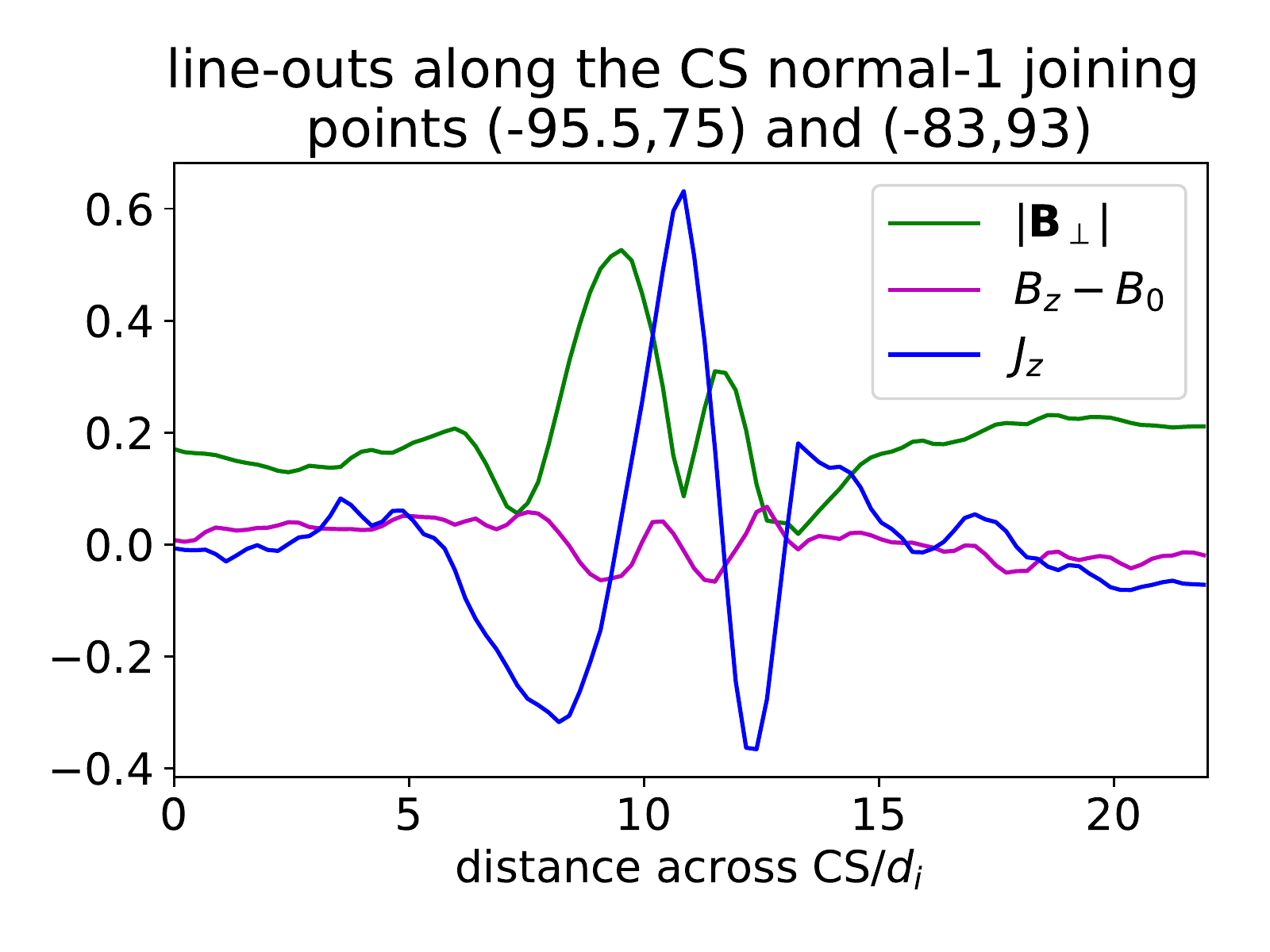}
    \includegraphics[clip=true,trim=0.25cm 0.7cm 0.25cm 2.1cm,width=0.32\linewidth]{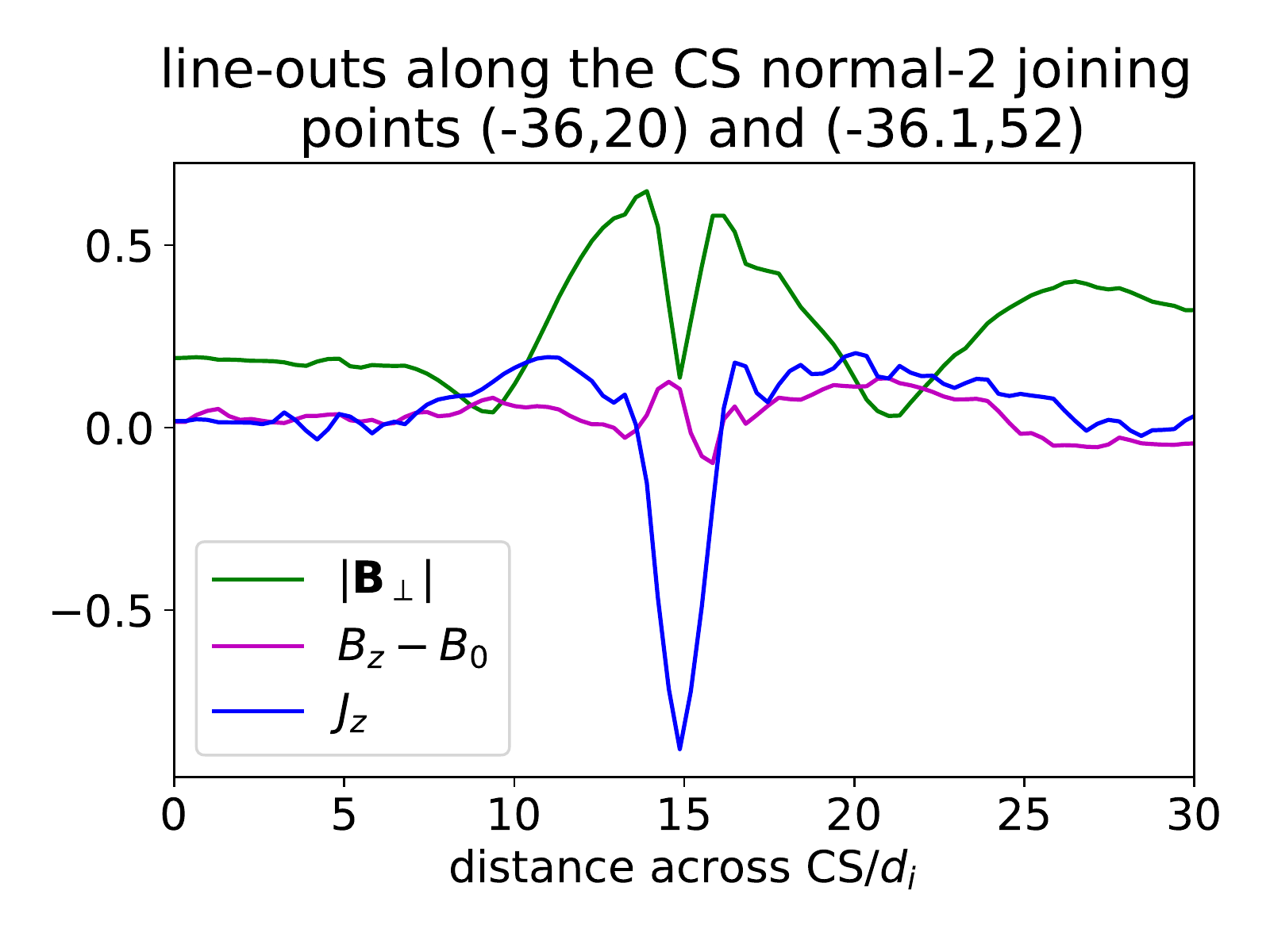}
    \includegraphics[clip=true,trim=0.25cm 0.7cm 0.25cm 2.1cm,width=0.32\linewidth]{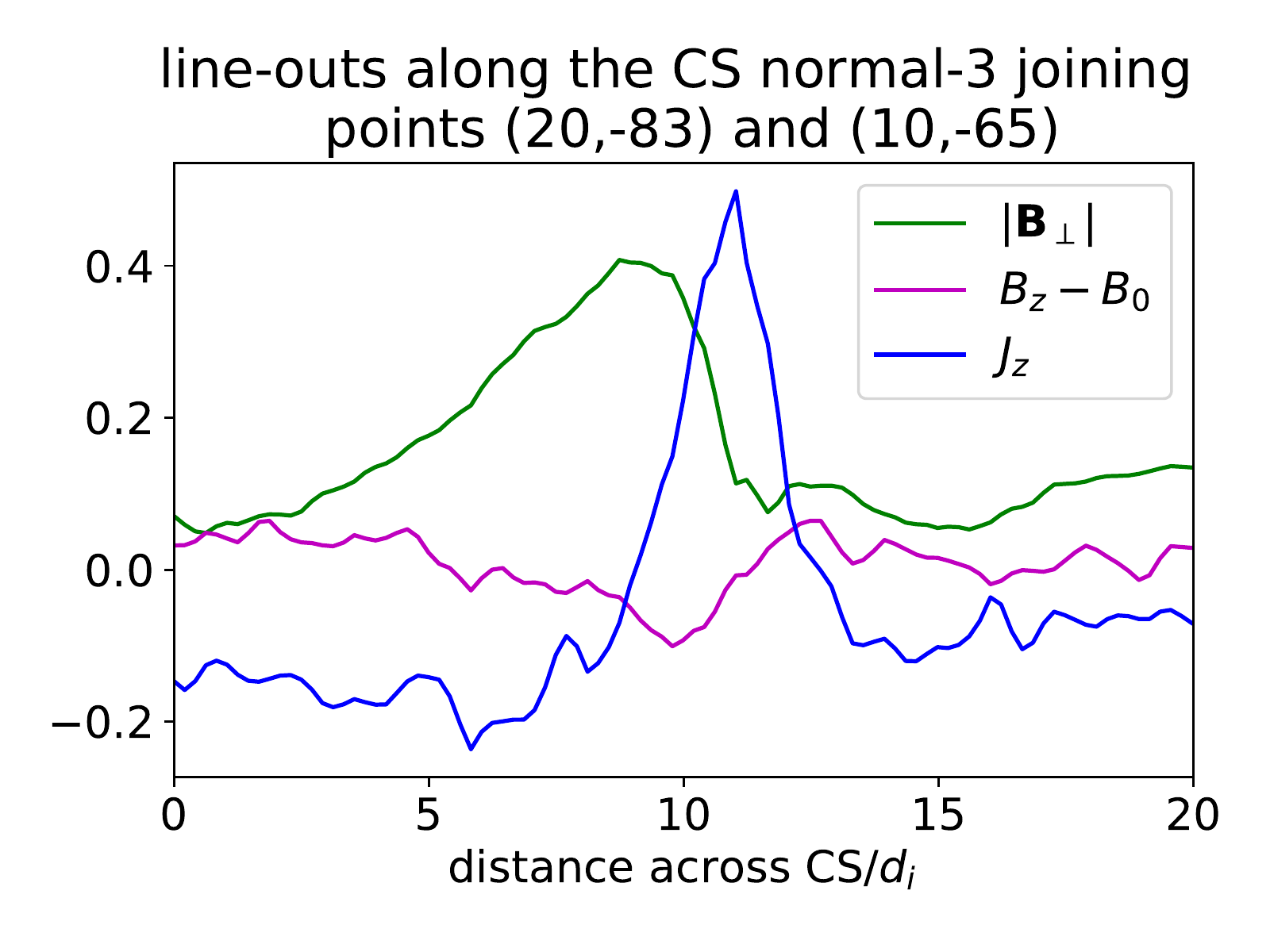}
        \caption{Line-outs of $|B_{\perp}|/B_0$, $(B_z-B_0)/B_0$ and $j_z/(n_0ev_{Ai})$ across the three CSs 
        (from left to right: CS-1, CS-2, CS-3) highlighted in Fig. \ref{fig:jz} at $\omega_{ci}t=50$.}
        \label{fig:lineouts_b}
    \end{center}
\end{figure}

In Sec.~\ref{sec:data}, we will use the jump condition of $u_e/u_i$  to detect CSs in solar wind by applying the condition to the time series measurements made by WIND spacecraft in quiet solar wind from 00:00, February 13 to 12:00, February 14, 1998 with a resolution of 3 seconds.  For these observations (solar wind speed $V \sim$ 400 Km/sec, density $\sim$ 10 cm$^{-3}$), the spacecraft travels approximately 1200 km $\approx 16.4 \,d_i$ during a 3 seconds measurement. Therefore, the time series measurements by the WIND spacecraft are effectively averaged over a distance of approximately 16 $d_i$.  We  average the simulation values of $u_e$ and $u_i$ over a distance of 16 $d_i$ along an assumed spacecraft trajectory passing the simulation plane at $y=-95.4 \, d_i$ to find out if $\bar{u}_e/\bar{u}_i$ still jumps at CSs ($\bar{u}_e$ and $\bar{u}_i$ are the averaged values) or the jump is washed out by the averaging procedure. Fig. \ref{fig:trajectory_lineouts_ueDui} shows the assumed trajectory and line-outs of $\bar{u}_e/\bar{u}_i-1$ and $j_z$ (without averaging) along the trajectory at $\omega_{ci}t$=50. Almost for every current sheet crossed by the assumed spacecraft trajectory, there is an associated jump of $\bar{u}_e/\bar{u}_i$, though, with a value below unity much smaller than the jump values of $u_e/u_i$ due to the averaging. The locations of the jumps are shifted with respect to the locations of corresponding current density peaks. Therefore, the jump condition of $\bar{u}_e/\bar{u}_i$ can be used to detect CSs in the solar wind observations made by spacecraft with a typical data resolution of one-three seconds.

\section{Theoretical estimate of the electron-to-ion bulk speed ratio}
\label{sec:theory}
For a quantitative comparison with observations it is appropriate to estimate the expected values of electron and ion bulk flow velocities. Theoretical estimates for the ratio of the out-of-plane electron and ion bulk velocities, $|u_{ez}|/|u_{iz}|$, was obtained approximating ion response as un-magnetized \cite{jain2020}. Here we estimate theoretically the ratio of the total electron and ion bulk velocities, $u_e/u_i$ under the approximation of un-magnetized ions.

For CSs with thicknesses of the order of ion gyro-radius $\rho_i=\sqrt{\beta}d_i$, ions can be approximated as un-magnetized while electrons are still tied to the magnetic field lines. Fig. \ref{fig:lineouts_b} shows lineouts across CSs CS1-CS3 of the turbulent magnetic field components perpendicular ($|\mathbf{B}_{\perp}|$) and parallel ($B_z-B_0$) to the applied magnetic field $B_0\hat{z}$. The turbulent magnetic field near the current sheet center, where current density peaks, is an order of magnitude smaller than the applied magnetic field ($|\mathbf{B_{\perp}}|/B_0 \sim (B_z-B_0)/B_0 \sim 0.1$).  Therefore, we take parallel and perpendicular directions inside CSs  (approximately) with respect to the applied magnetic field $\mathbf{B}_0=B_0\hat{z}$.  
We can then obtain ion bulk velocity $\mathbf{u}_i$ from ion's momentum equation neglecting Lorentz force, perpendicular electron bulk velocity $\mathbf{u}_{e\perp}$ as $E \times B$ drift from Ohm's law and parallel electron bulk velocity $\mathbf{u}_{ez}$ from Ampere's law.
 \begin{eqnarray}
   \frac{\partial\mathbf{u}_i}{\partial t} &=&\frac{e\mathbf{E}}{m_i}
   \label{eq:ion_mom}\\
   \mathbf{u}_{e\perp}&=&\frac{\mathbf{E}_{\perp}\times \mathbf{B}}{B^2}\label{eq:ueperp}\\
   \mathbf{u}_{ez}&=&\mathbf{u}_{iz}-\frac{\nabla_{\perp} \times \mathbf{B}_{\perp}}{\mu_0 n e}\label{eq:ampere}
 \end{eqnarray}
 
 Electric and magnetic fields are related by Faraday's law.
\begin{eqnarray}
   \nabla_{\perp}\times \mathbf{E}=-\frac{\partial \mathbf{B}}{\partial t}\label{eq:faraday}
 \end{eqnarray}

Here $\nabla_{\perp} \equiv \hat{x}\partial/\partial x+\hat{y} \partial/\partial y$. In Eq. \ref{eq:ion_mom}, the convective derivative $(\mathbf{u}_i.\nabla) \mathbf{u}_i$ is neglected compared to the time derivative  inside CSs   under the approximation $|(\mathbf{u}_{i}.\nabla)\mathbf{u}_i| / |\partial \mathbf{u}_{i}/\partial t| \sim u_{i,\perp}/v_{Ai} \sim 0.1 << 1$ (for $\partial/\partial t \sim v_{Ai}/L$ and $ \nabla_{\perp} \sim L^{-1}$) as it was demonstrated by simulations \cite{jain2020}.
Eqs. \ref{eq:ion_mom} and \ref{eq:ueperp} give estimates as $u_{i\perp}\sim L E_{\perp}/d_iB$, $u_{iz} \sim  LE_{z}/d_iB$ and 
   $u_{e\perp} \sim E_{\perp}B_z/B^2$.  
   
 Estimating $E_{\perp} \sim b_z v_{Ai}$ and $E_z \sim b_{\perp} v_{Ai}$  from Faraday's law, we get, 
\begin{eqnarray}
  u_{i\perp}&\sim&  v_{Ai} L b_z/ d_iB \label{eq:uiperp}\\
  u_{iz}&\sim&  v_{Ai} L b_{\perp}/d_iB \label{eq:uiz}\\
  u_{e\perp}&\sim &v_{Ai} b_zB_z/B^2\label{eq:ueperp1}
\end{eqnarray}

Here $b_{\perp}=|\mathbf{B}_{\perp}|$ and $b_z=B_z-B_0$ are turbulent magnetic field components. The first term ($|u_{iz}| \sim v_{Ai}L b_{\perp}/d_iB$) on the RHS of Eq. \ref{eq:ampere} can be neglected in comparison to the second term ($|\nabla_{\perp}\times \mathbf{B}_{\perp}|/\mu_0 ne \sim v_{Ai}d_ib_{\perp}/L B$) for perpendicular spatial scale lengths $L << d_i$ giving,
\begin{eqnarray}
  u_{ez}&\sim& v_{Ai}d_ib_{\perp}/L B \label{eq:uez}.
\end{eqnarray}
The parallel and perpendicular components of electron and ion bulk velocities can now be compared inside CSs using Eqs. (\ref{eq:uiperp})-(\ref{eq:uez}). From Eqs. (\ref{eq:uiperp}) and (\ref{eq:ueperp1}), $u_{e\perp}/u_{i\perp}\sim (B_z/B) (d_i/L)$. For un-magnetized ions ($L < \rho_i \sim d_i$) and $B_z\sim B \sim B_0$, $u_{e\perp}/u_{i\perp}\sim d_i/L > 1$ consistent with the results of the hybrid simulations \cite{jain2020}. Inside CSs, the perpendicular ion bulk velocity is typically smaller than the perpendicular electron bulk velocity due to the demagnetization of ions. From Eqs. (\ref{eq:uiz}) and (\ref{eq:uez}), $u_{ez}/u_{iz} \sim d_i^2/L^2 > 1$. Note that both the ratios $u_{e\perp}/u_{i\perp}$ and $u_{ez}/u_{iz}$ are greater than unity but $u_{ez}/u_{iz} > u_{e\perp}/u_{i\perp}$ consistent with the simulations \cite{jain2020}. Using $B_z\sim B$, $b_{\perp} \sim b_z$ and $d_i^2/L^2 >> 1$, the ratio of the net electron and ion bulk velocities, $u_e=(u_{e\perp}^2+u_{ez}^2)^{1/2}$ and $u_i=(u_{i\perp}^2+u_{iz}^2)^{1/2}$ respectively, can be written as, 
\begin{eqnarray}
  \frac{u_e}{u_i}&\sim& \frac{d_i^2}{L^2} \frac{b_{\perp}}{(b_{\perp}^2+b_z^2)^{1/2}}.
\end{eqnarray}

The ratio $u_e/u_i$ is smaller than the ratio $u_{ez}/u_{iz}$ by a factor of the order of unity, again consistent with the results of the hybrid simulations \cite{jain2020}.
With the CS thinning, therefore, the current in the sheet is confirmed
to be increasingly carried by the electrons. 

\begin{figure}
  \begin{center}
    	\includegraphics[width=0.8\linewidth]{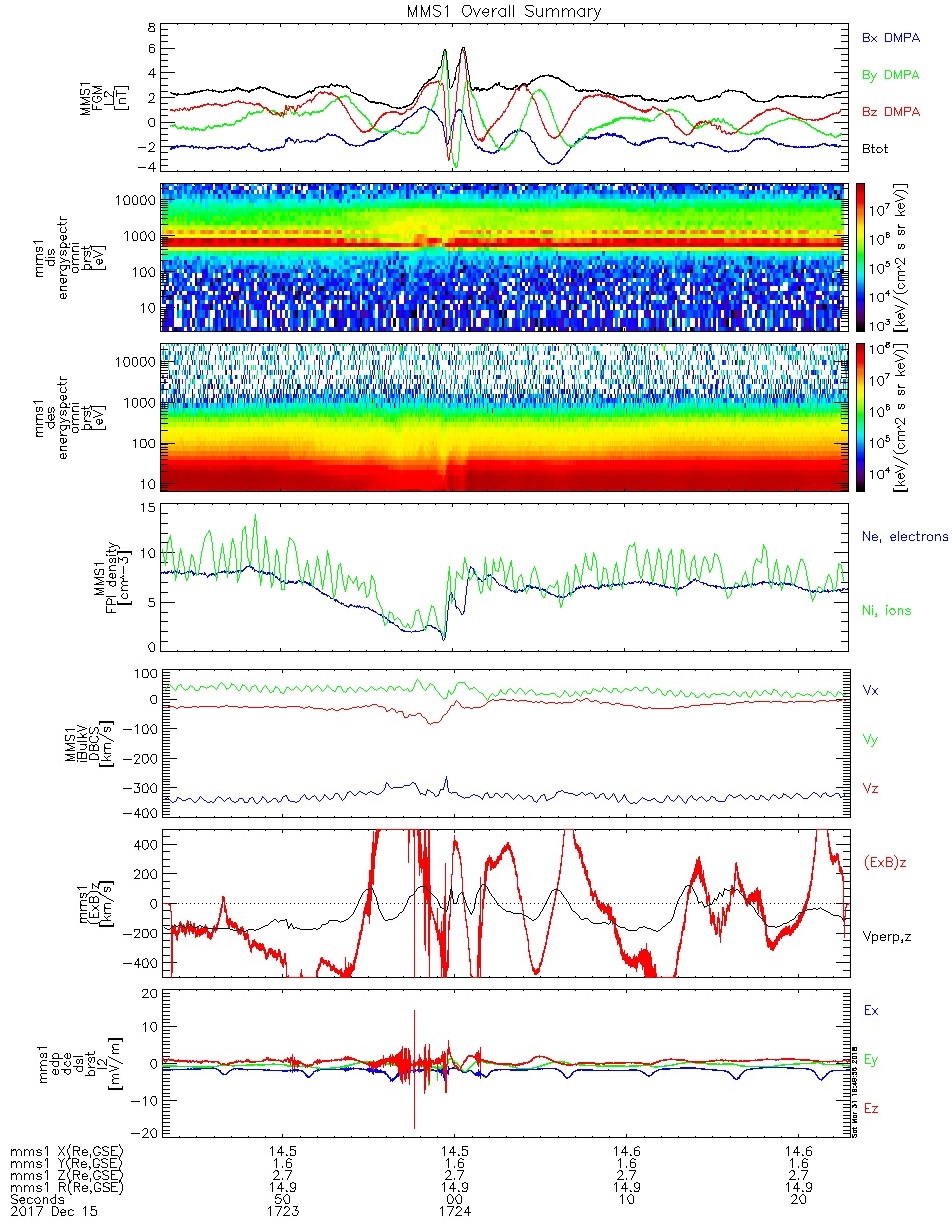}
	\caption{Bifurcated current sheet formed by ion- and electron-dominated currents as observed by the MMS1 satellite in the solar wind. From top to bottom: total IMF and three IMF components in the GSE coordinate system; ion and electron energy flux spectrograms; ion and electron densities; ion velocity components; ion drift speed in Z direction; electric field vector. Temporal resolution is 150 milliseconds for ions and 30 milliseconds for electrons.
}
	\label{fig:MMS1}
    \end{center}
\end{figure}

\section{
High- and low-resolution observations of electron-dominated current sheets in the solar wind}
\label{sec:data}

As noted in the Introduction, in most cases one cannot directly observe electron-scale CSs in the solar wind, first of all, because of the absence of the constellation-type spacecraft operating there. Finding the location of CSs of any type in the solar wind is always a matter of the analysis of several parameters that specifically vary at CS crossings, as shown in ~\citep{khabarova2021, Khabarova2021SSRv}. 
The second problem is that a resolution of measurements of the magnetic field in the solar wind is usually about one second that corresponds to one-two proton gyroradii. This is satisfactory for identifying ordinary ion-scale CSs, but this makes impossible direct observations of much thinner electron-scale CSs. As for electron-dominated CSs in the solar wind, so far there have not been studies concerning their properties and identification.

 Meanwhile, it is useful to know at least an approximate location of electron-dominated CSs because they may carry currents larger than those carried by ions ~\citep{Podesta2017SoPh, Wang_et_al2018}.
We will show that despite the obstacles discussed above, one may consider indirect observational signatures of electron-dominated CSs to recognize them from the solar wind data. A case study below confirms that at least some electron-dominated CSs can be characterized by very intense currents comparable by width with well-known ion-dominated CSs. 

First, we show an example of the electron-dominated CS observed by MMS with an unprecedentedly high resolution (150 milliseconds for ions and 30 milliseconds for electrons), checking the hypothesis that its impact on plasma parameters is strong and spatially wide enough to be detected with far lower resolution.  
Second, we apply the $u_e/u_i$ signature of electron-dominated CSs to the solar wind data on the electron and proton velocity at 1 AU obtained from the Wind spacecraft. We compare locations of electron-dominated CSs found via sharp variations of $u_e/u_i$ with locations of ion-dominated CSs identified with the method described by ~\citep{khabarova2021}.
We  claim below that a jump in the $u_e/u_i$ parameter can potentially point out a strong electron-dominated CS located somewhere within the region crossed by a spacecraft for 3 seconds (which is a typical temporal resolution of the solar wind spacecraft for the magnetic field). 

\begin{figure}
  \begin{center}
    	\includegraphics[width=0.5\linewidth]{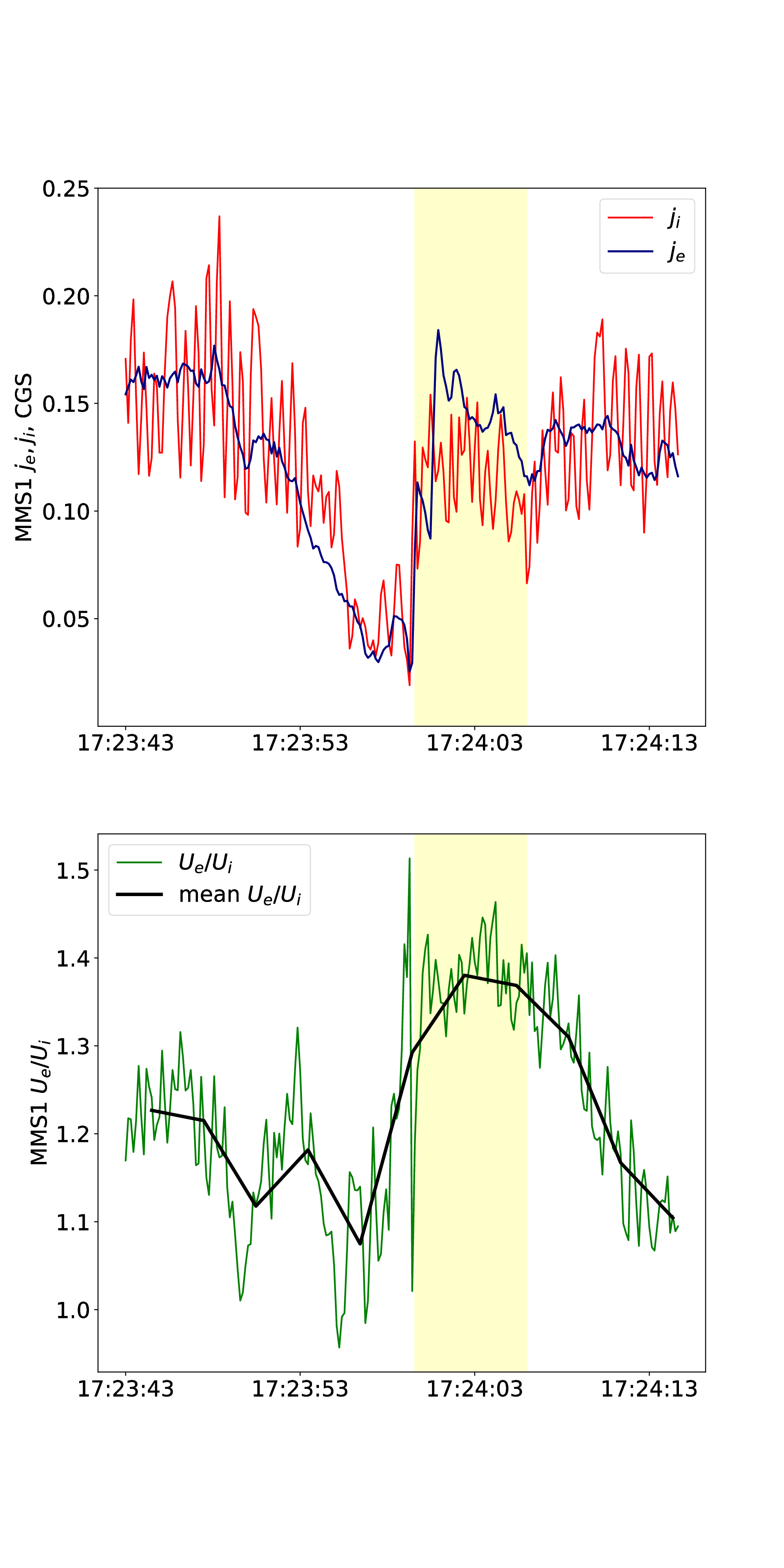}
	\caption{Electric current densities and the $u_e/u_i$ parameter observed by MMS1 through the CS shown in Fig.~\ref{fig:MMS1}. From top to bottom: ion current density (red) and electron current density (blue); the $u_e/u_i$ parameter calculated with the 150 millisecond resolution (green line) and its 3 second average (black line). Yellow stripe indicates the current and plasma variations associated with the electron-dominated CS. Observations confirm that the effect of $u_e/u_i$ increasing at electron-dominated CSs is strong enough not to disappear when plasma measurements are analyzed with a resolution typical for the solar wind spacecraft.
}
	\label{fig:MMS2}
    \end{center}
\end{figure}

\subsection{Example of 
high-resolution} observations of the electron current dominated over the ion current as detected by MMS in the solar wind
In the previous, theoretical section we showed that a strong electron-dominated CS crossed by a spacecraft in the solar wind can potentially be spotted even with a rough one-three-second resolution because of the impact of a strong electric current on the plasma. The latter can be visible via the $u_e/u_i$ ratio increase. To illustrate the same effect with in situ spacecraft measurements, we use the MMS mission data since its resolution allows studying the fine structure of CSs of all types, including the thinnest electron-scale CSs  ~\citep{burch2016, Leonenko_et_al2021}. 
The highest resolution data are avalable when the burst mode is triggered. One can find more information about the burst mode measurements in ~\citep{Argall2020}. Burst mode plots and data can be accessed on the mission website: \url{https://lasp.colorado.edu/mms/sdc/public}.

Fig.~\ref{fig:MMS1} shows MMS1 observations performed in the burst mode from 17:23:43 to 17:24:23 UT on 15 December 2017 when MMS was in the solar wind. This is an overview of observations for this period provided on the mission website \url{https://lasp.colorado.edu/mms/sdc/public/data/sdc/burst/all_mms1_summ/2017/12/15/burst_all_mms1_summ_20171215_172343.png} . Sharp variations in $B_z$ and $B_y$ interplanetary magnetic field (IMF) components crossing zero lines coincide with the corresponding variations and enhancements in the ion flux in the keV range and electron flux up to hundreds eV. The amplitude of the electric field increases considerably together with sharp changes in the drift velocity. Such plasma objects can be classified as bifurcated CSs. An example of these CSs is shown in Fig.~\ref{fig:jz} (see the pairs of blue and red curves). Sometimes they are also treated as magnetic holes or crossings of elongated flux tubes with borders representing CSs if one considers possible counterparts in 3D. What is important for the particular study is that if one calculates the current density for ions and electrons, then it becomes clear that the ion current dominates at one side of the dip in the total magnetic field and the electron current exceeds the ion current at the other side. 
This can be seen in the upper panel of Fig.~\ref{fig:MMS2}. The electron current density is shown by blue and the red curve represents the ion current density. The time period during which the electron current dominates is highlighted by the yellow stripe. The corresponding increase of the $u_e/u_i$ ratio is seen in the lower panel of Fig.~\ref{fig:MMS2}. One can find that changes in the plasma parameter are clearly pronounced at the crossing of the electron-dominated CS, they are smoother and even wider than the corresponding changes in the electric current density in the upper panel. The effect does not disappear under the three second averaging, which means that using ordinary data from spacecraft like Wind may help reveal strong electron-dominated CSs in the solar wind.

\subsection{Locating electron-dominated current sheets by means of the electron-to-ion bulk speed ratio 
from the low-resolution WIND spacecraft data} 
\label{sec:results}

{We use below data from Wind, a spacecraft operating at 1 AU in the solar wind, at the 1st Lagrangian point. It has a typical resolution in terms of plasma and magnetic field measurements in the solar wind, far lower than a data resolution of such magnetospheric missions as MMS or Cluster, but at the same time, this is one of rare spacecrat that allows measuring the electron velocity.} Solar Wind Experiment (SWE) Electron Data Sources are available at NASA’s Space Physics 
Data Facility (SPDF) HTTPS site \url{https://cdaweb.gsfc.nasa.gov/index.html}.
They allow finding the velocity of electrons ($u_e$) necessary for the study.
We further use the ion (proton) velocity ($u_i$~) obtained by the Wind spacecraft 
3-D Plasma and Energetic Particle Investigation experiment 
(Wind 3DP,  \url{http://sprg.ssl.berkeley.edu/wind3dp}).
From those data we calculate the $u_e/u_i$~ ratio and find the total magnetic field $B$. 

Additionally, we employ the solar wind key parameters to compile a list of ion current 
sheets via the automated method that considers sharp variations in the total magnetic field, $\beta$~, and 
the Alfv\'en speed $V_a$~ to the solar wind speed $V$~ ratio~\citep{khabarova2021}. 
This is the basis of the three-parameter method, using which the IZMIRAN database of 
CSs has been built (see \url{https://csdb.izmiran.ru/}).  
Summarizing, the following Wind data from the SPDF website have been used:

- WI\underline{ }H2\underline{ }MFI - Wind Magnetic Fields Investigation, high-resolution definitive data (IMF);

- WI\underline{ }PM\underline{ }3DP - Ion moments (the velocity, the density, and the temperature of the solar wind protons);

- WI\underline{ }EM\underline{ }3DP - Electron Plasma moments (the electron velocity).

The $u_e/u_i$~ data have the three second resolution, and the three key parameters to identify the ion current sheet location via the method described in ~\citep{khabarova2021} are calculated with a one second cadence. The noise effects can be neglected after the procedure of setting the threshold for spotting only strong CSs (see below). 

\begin{figure}
  \begin{center}
    	\includegraphics[width=0.9\linewidth]{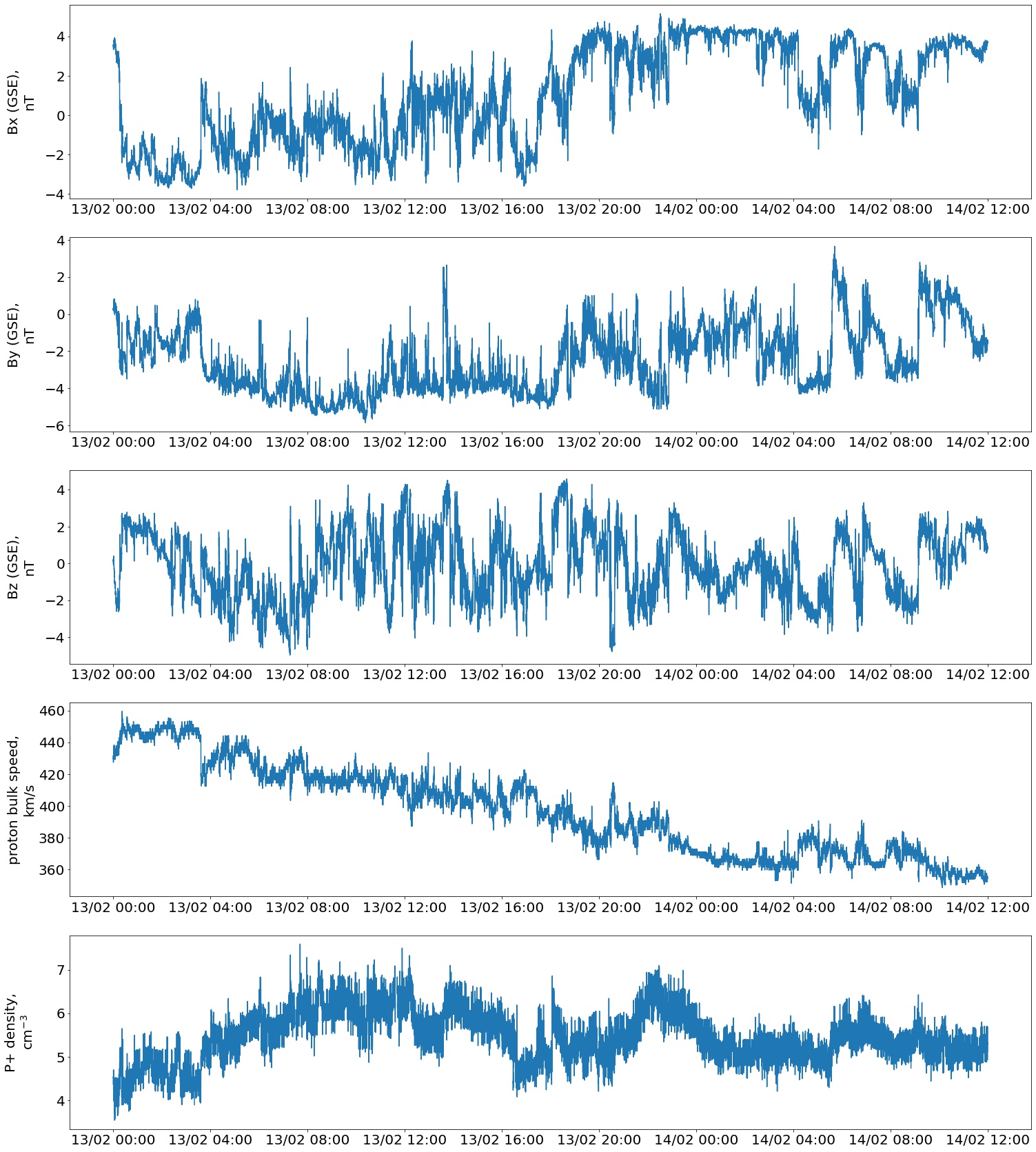}
	\caption{A period of a  quiet solar wind: from 00:00 on February 13, 1998 to 12:00 on
	February 14, 1998. 
	From top to bottom: 
	the three components of the IMF in the GSE coordinate system, 
	the proton bulk speed $(u_i)$, 
	and the proton density $(p+)$ 
	as observed by the Wind spacecraft.}
	\label{fig:FigureX1}
    \end{center}
\end{figure}

We have selected a very quiet solar wind period from 00:00 February 13, 1998 to 12:00 February 14, 1998 during which the near-Earth plasma was not affected by either interplanetary coronal mass ejections (ICMEs) or stream interaction regions (SIRs). One can see in the three upper panels of Fig.~\ref{fig:FigureX1} that the $B_y$~ and $B_z$~ components of the IMF in the Geocentric Solar Ecliptic (GSE) coordinate system vary around zero, and the $B_x$~ component shows a slow transition from the negative to positive IMF sector, suggesting a crossing of the heliospheric plasma sheet (HPS). The HPS is a wide area filled with numerous CSs produced, on the one hand, by magnetic reconnection and instabilities developing at the HCS embedded in the HPS, and, on the other hand, by the same processes occurring at other strong and long-lived CSs representing an extension of former streamers expanding from the solar corona ~\citep{Malova_et_al2020}. Because of this, the IMF components may vary around zero for hours, and the IMF does not immediately change its direction at the HCS within the HPS ~\citep{Khabarova2021SSRv}. 

As seen in the two bottom panels of Fig.~\ref{fig:FigureX1}, the spacecraft is in the slow solar wind with the ordinary, not elevated solar wind density. The proton bulk speed is lower than 450 km/s and the proton density curve lies below the level of 10 particles per $cm^3$. Therefore, the interval is ideal for the exploration of turbulence enhanced by products of magnetic reconnection at CSs within and in the vicinity of the HPS.

\begin{figure}
  \begin{center}
    	\includegraphics[width=0.8\linewidth]{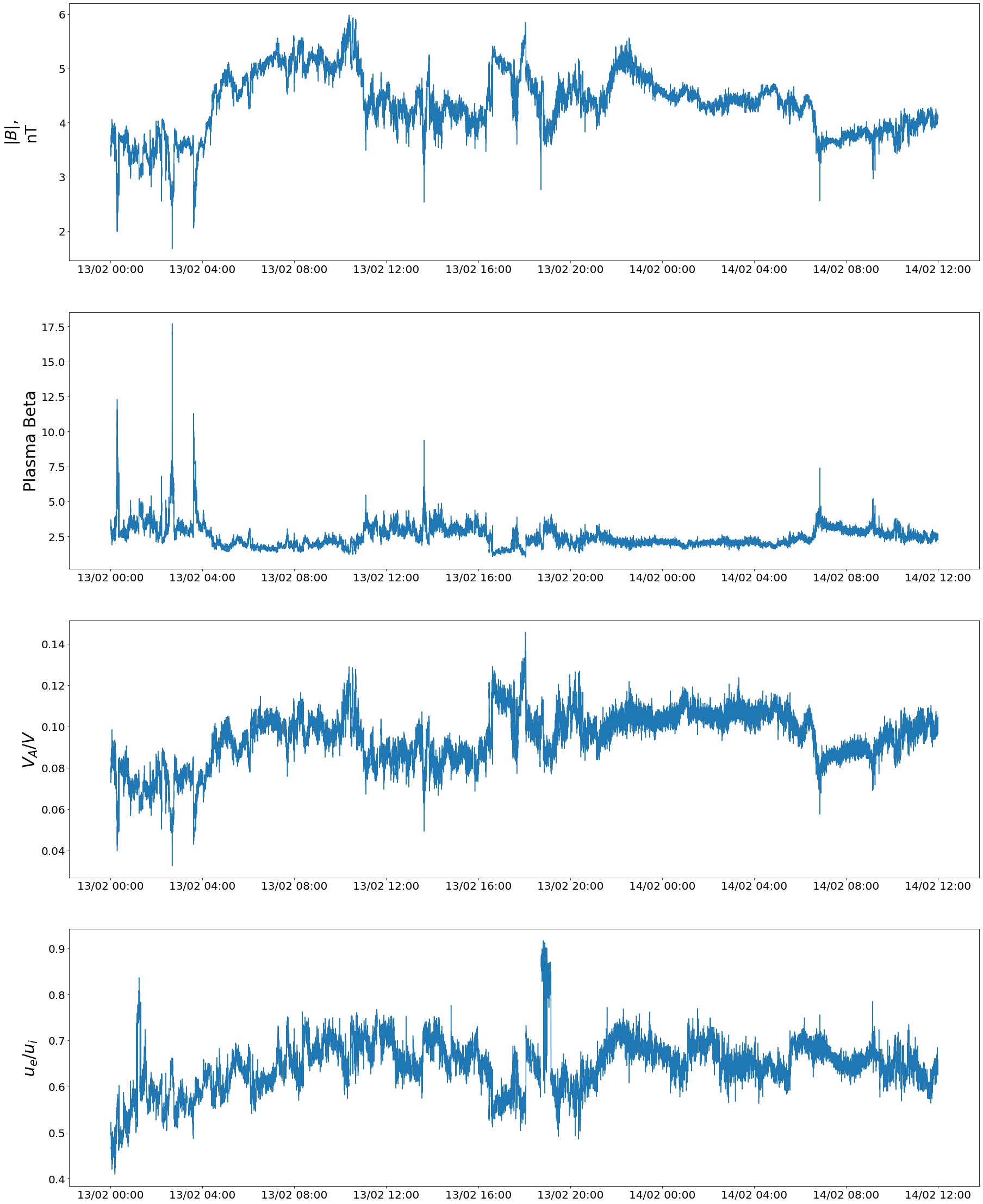}
	\caption{Key parameters helping identify both electron- and ion-dominated CSs for the same period as shown in Fig.~\ref{fig:FigureX1}. 
	From top to bottom: the IMF strength $B$, $\beta$, $V_A/V$, and $u_e/u_i$.}
	\label{fig:FigureX2}
    \end{center}
\end{figure}

\begin{figure}
  \begin{center}
    	\includegraphics[width=0.8\linewidth]{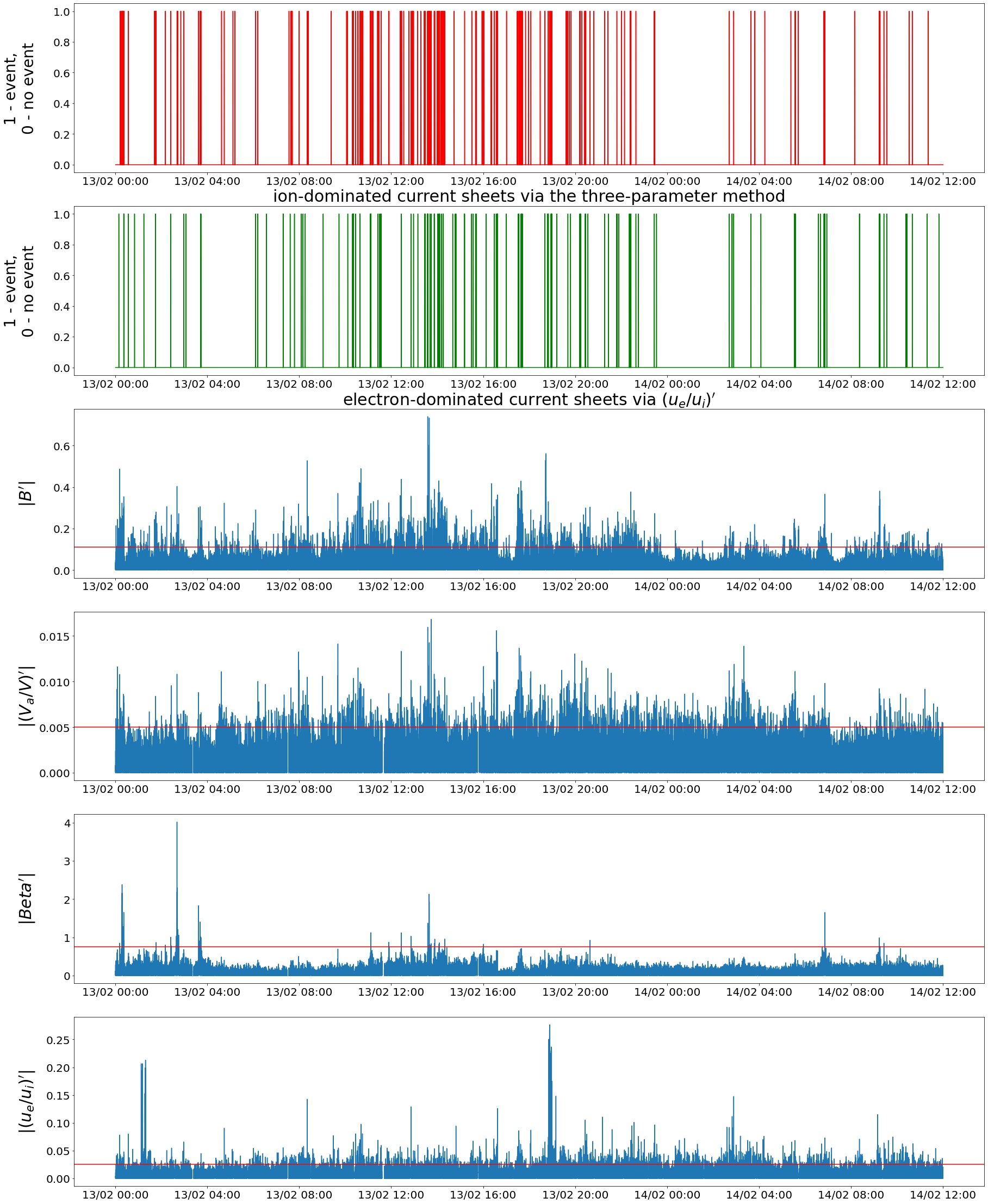}
	\caption{Location of ion- and electron-dominated  CSs identified for the same period as shown in Fig.~\ref{fig:FigureX1}. 
	From top to bottom: Location of ion-dominated CSs identified via the three-parameter method ~\citep{khabarova2021} (red lines) versus the location of electron-dominated CSs found via	 $(u_e/u_i)'$ and $B'$ (green lines), respectively.
	Module of derivatives of $B$, $V_A/V$, $\beta$, and $u_e/u_i$ used to show the sharpest changes in the parameters indicating CS crossings.} 
	\label{fig:FigureX3}
    \end{center}
\end{figure}

First, we create a list of ion CSs, following ~\citep{khabarova2021}.
  At the next step we identify plasma structures presumably associated with electron-dominated  CSs as predicted by the simulations discussed above. Such structures are supposed to be characterized by sharp variations in $u_e/u_i$ and simultaneous variations in the IMF module $B$. Therefore, we identify them by calculating derivatives of $u_e/u_i$ and $B$ and setting up the noise thresholds as discussed below. Then we compare both rows to find similarities and differences.

Fig.~\ref{fig:FigureX2} shows variations of the IMF strength in the upper panel, according to which one may approximately estimate how often and where the location of the strongest ion-dominated CSs are crossed by the Wind spacecraft. Sharp dips in $B$ seen with a one-second resolution correspond to crossings of neutral lines at CSs when at least one of the IMF components equals zero in the corresponding reference system. This is the simplest way to identify CSs by eye known among observers. Additionally, $\beta$ and $Va/V$ variations are taken into account since statistics shows that the plasma beta jumps and the $Va/V$ ratio falls at ion CSs. To make the changes more pronounced, derivatives of these parameters are taken to identify CSs as described in ~\citep{khabarova2021}.

 $u_e/u_i$ is below 1 in the lower panel of Fig.~\ref{fig:FigureX2} despite the theory predictions of a $u_e/u_i$ jump above 1 at electron-dominated CSs (see above). This is a result of working with real observations and necessary data averaging. The point is that this parameter is always below 1 in the background plasma around such CSs. CSs in which electrons carry the main electric current are thin in comparison with the space surrounding them and included into averaging. Predictions show that the visible input of the current into the $u_e/u_i$ jump should last for about one second. The input of the parameter exceeding 1 into the 3-second-averaged picture is low in the background of dominating $u_e/u_i$ values far below 1. Only the strongest electron-dominated CSs can be detected with the 3-second resolution and seen as sharp increases of the $u_e/u_i$ parameter in Fig.~\ref{fig:FigureX2}.  

The result of the identification of ion CSs via the three-parameter method is shown in the first panel of Fig.~\ref{fig:FigureX3} in the form of red bars. Zero means no CS and one corresponds to the presence of a CS identified with the application of the following thresholds that cut off the noise: $B' \le -0.11$; $V_a/V \le-0.005$; and $\beta \ge 0.75$ . The thresholds are shown as red horizontal lines. Imposing the thresholds help us neglect possible device noise and too weak CSs to be of interest. The other panels show variations in the parameters that help detect CSs with an automated method, running the corresponding code similar to that described in ~\citep{khabarova2021} . 

We have found that, analogous to the $Va/V$ parameter used in ~\citep{khabarova2021}, $u_e/u_i$ itself displays the location of CSs worse than its derivative (compare the corresponding panels in Fig.~\ref{fig:FigureX2} and Fig.~\ref{fig:FigureX3}).
The second (green) panel in Fig.~\ref{fig:FigureX3} shows the location of electron CSs identified using the proposed $u_e/u_i$ parameter and $B'$ . The corresponding noise-cutting threshold is $(u_e/u_i)' \ge 0.05$. The other panels show variations in the parameters that help detect CSs with an automated method, running the corresponding code similar to that described in ~\citep{khabarova2021} . Although the exact location of ion CSs (red) and electron CSs (green) does not always coincide, the both red and green panels show clear clustering of CSs in the same places, and the strongest CSs easily visible as the sharp $B$ decreases and the plasma beta jumps are successfully identified via both $(u_e/u_i)'$ variations and the three-parameter method. 

Some difference between the location of ion and electron CSs can be explained, first, by the fact that the $(u_e/u_i)'$ parameter often catches an inner thin CS with the current produced by elections, which is embedded in the wider "ion" current sheet. Purely techniclaly, two methods having different accuracies always return a little different location of the corresponding structures. Second, the $(u_e/u_i)'$ parameter is supposed to be more sensitive to thin CSs born as a result of pure turbulence than to CSs produced by magnetic reconnection at strong large-scale CSs such as the HCS (see the Introduction).

Fig.~\ref{fig:FigureX3} shows that despite very similar clustering, electron CSs may be observed without any association with ion CSs, and vice versa. This is an interesting result because this may reflect not just a different sensitivity of different methods but have a certain physical sense, allowing us to suggest that CSs of very different types can form under different conditions. However, a confirmation of this idea requires thorough investigations of properties of electron and ion CSs observed in the differently-originated solar wind flows or streams. 

Therefore, preliminary results support the idea that the electron to proton velocity ratio can be considered as one of key parameters to detect electron-dominated CSs. Further studies will show details of how ion and electron-dominated CSs are related and why they sometimes exist separately.

\section{Conclusions and discussion} 
\label{sec:conclusion}

This study suggests a way to identify the strongest electron-dominated CSs in the solar wind via an automated method that may be used for statistical purposes. Electron-dominated CSs are current layers of various origins in which the electron current exceeds the ion current. Electron-scale CSs are a sub-set of electron-dominated CSs that can also be of ion scales. The main idea is based on theoretical and observational findings that the most intense electron currents impact the plasma significantly and can be spotted in the solar wind at the scales of thousands kilometers.

No electron-dominated CSs have been observed in the solar wind before because it was supposed that they should have an electron-scale width unresolvable by spacecraft. This study shows the first example of a CS with the current driven by electrons as observed by the MMS mission in the solar wind, outside the magnetoshere and the foreshock area. It is found that electron-dominated CSs can be much wider that expected before, and variations in the electron and ion speeds can reflect the occurrence of such CSs at the scales up to several ion gyroradii. Observations show that the impact of electron-dominated CSs on the surrpounding plasma is significant. It can be seen in the vicinity of +/- several CS widths with respect to the location of the particular CS.  Further studies with the help of the MMS mission are needed to analyze key properties and stability characteristics of electron-dominated CSs in the solar wind. 

Beginning with the first investigations of thin magnetospheric CSs based on the unique MMS data, it has been known that electrons carry the strongest electric current in CSs. Only electrons are significantly heated and move fast, while ions keep the average temperature and display almost no acceleration  ~\citep{Wang_et_al2018, Macek2019, Bandyopadhyay2021}. Recent numerical simulations show that, indeed, electron-dominated CSs are associated with an increase of the 
electron-to-ion bulk speed ratio $u_e/u_i$ and with electrons 
becoming the main carriers of the electric current. Numerical studies of the CS formation in turbulent plasmas
by fully kinetic as well as by hybrid code simulations (in which ions are
considered as particles and electrons as a fluid) found that thin electron-scale 
CSs can be formed inside ion-scale thicker CSs (e.g., ~\citep{azizabadi2021,Malova_et_al2017}). 
Simulations done for the magnetospheric conditions suggest that electron-dominated CSs can also exist independently
of ion CSs (e.g., see ~\citep{Zelenyi_et_al2020} and references therein). 

High-accuracy observations of magneto-plasma structures at the magnetopause and in the tail of the terrestrial magnetosphere as well as in planetary magnetospheres show that electron current layers are usually found at cites where CSs become significantly thinned and ready for reconnection, or when magnetic reconnection is already underway ~\citep{Nakamura2006, Panov2006, Runov2008, Grigorenko_et_al2019, Zelenyi_et_al2020, Hubbert_et_al2021}. 
A thickness of such electron-dominated CSs may be as small as a few gyroradii of thermal electrons  ~\citep{Leonenko_et_al2021}. 
Most of them are embedded in wider ion CSs, but single electron CSs can be observed too ~\citep{Wang_et_al2018, Bandyopadhyay2021} .
Although electron CSs possess very similar characteristics in different plasmas, their lifetime and stability characteristics are different ~\citep{Zelenyi_et_al2008, Zelenyi_et_al2010, Zelenyi_et_al2019}. It seems that electron CSs not associated with ion CSs are ubiquitous in the Martian magnetosphere but rather rare in the Earth’s magnetotail.

Before this study, the following question has remained opened: whether it is possible to find signatures of electron-dominated-CSs 
in the solar wind plasma? Existing methods of identifying CSs in the heliosphere are focused on ion-dominated CSs, mainly considering the magnetic field behaviour and, rarely, 
the behaviour of plasma parameters ~\citep{khabarova2021}. There have not been comprehensive studies of electron-dominated CSs in the solar wind, for many technical reasons. Their automated identification and statistical investigations have been thought impossible for a long time. Even finding an approximate location of electron CSs is a difficult task, and case studies employing magnetospheric missions in the solar wind for this aim are extremely rare (e.g., ~\citep{Mistry_etal2015}).  Meanwhile, studying such CSs is especially important because simulations show that electron CSs most probably carry the largest electric currents in the solar wind ~\citep{Podesta2017SoPh}.    

To solve this problem at least partially we suggested to use indirect signatures of electron CSs based on the results of numerical simulations. In this study we applied the $u_e/u_i$ criterion of the existence of an electron-dominated CS visible even at ion scales to the solar wind at 1 AU utilizing the Wind spacecraft data. We selected a quiet solar wind period within which numerous sharp variations 
of $u_e/u_i$ were observed, suggesting that the most pronounced changes of $(u_e/u_i)'$ in a combination with those of $B'$ may point out an approximate location of strong electron CSs, as both theoretical predictions and in situ observations show. 

Then the location of electron CSs identified that way was compared with the location of ion CSs identified via the other method ~\citep{khabarova2021}. It was found that the structures presumably indicating electron-dominated CSs were mostly formed 
at/ or in the vicinity of ion-dominated CSs, showing the same clustering. An interesting point is that some of electron- and ion 
CSs were registered separately, without CSs of the other type found nearby. 

Summarizing, we report important properties of CSs formed
in the turbulent solar wind which are associated with electrons 
becoming the main current carriers. 
We conclude that, first, electron-dominated CSs in the solar wind can be of ion scales, and, second, the electron to proton velocity ratio may be considered 
as the major parameter identifying strong CSs of this type and allowing an analysis of their properties in turbulent plasmas. Based on MMS observations and simulations, we suppose that only the strongest electron-dominated CSs can be identified in the solar wind via such a method from observations made with a one-three second resolution typical for most spacecraft.

The results testing the hypothesis of the importance of the $u_e/u_i$ ratio in pointing to electron CSs are preliminary because electron CSs may be identified with a high degree of certainty only using several parameters, analogous to ion CSs, and with a high resolution. Here, we just discuss an important feature of the solar wind plasma which may be associated with thin CSs produced by electron currents. Future case studies employing data from the MMS mission will show what parameters are most important to find electron CS crossings and how to build a reliable method of electron CS identifying to investigate their properties statistically. So far, the proposed $u_e/u_i$ method can be considered as potentially useful for studies of turbulence in the solar wind and probing CSs in space plasmas.

\begin{acknowledgments}
\section*{Acknowledgments}
{The numerical simulations were supported by the German Science Foundation (DFG), 
project JA 2680-2-1. Part of the simulations were carried out on the HPC-Cluster of the 
Institute for Mathematics at the TU Berlin and the Max-Planck-Institute for Solar System 
Research G\"ottingen.
The work is further supported by the German-Russian project DFG-RSF BU 777-16-1. The observational part of the study is supported by the Russian Science Foundation grant No. 20-42-04418 (contributors: Olga Khabarova, Helmi Malova, and Roman Kislov). 
We acknowledge using the following open-access data published via the NASA’s SPDF 
website 
\url{https://cdaweb.gsfc.nasa.gov/index.html}: the magnetic field high-resolution definitive data (with many thanks to A. Koval, UMBC, NASA/GSFC) and the data on the velocity, the density, and the temperature 
of solar wind protons and electrons (with a gratitude to R. Lin and S. Bale, 
UC Berkeley) from the Wind spacecraft.
MMS data are obtained from the MMS Data Center (\url{https://lasp.colorado.edu/mms/sdc/public/}). Thanks to J. Burch, M. Kessel, B. Giles, and G. Le.
We acknowledge support of Open Access Publications by the German Research Foundation and by
the Open Access Publication Fund of TU Berlin.}
\end{acknowledgments}


\bibliography{references}{}
\bibliographystyle{aasjournal}
\end{document}